\documentclass[12pt]{article}
\usepackage{a4}
\usepackage{epsf}

\def\parag            {\vspace{2ex}}

\def\ul               {\underline }

\def\nn               {\nonumber}

\def\ma[#1,#2,#3,#4]  {{\left( \matrix{ #1  & #2 \cr
                                        #3  & #4 \cr } \right)}}

\begin{document}

\title{Topological sectors and the pion mass}

\author{
S. Elser 
\thanks{ 
supported by DFG research grant No.  WO 389/3-2; 
email: \mbox{elser@linde.physik.hu-berlin.de}
}
$\;$ and B. Bunk
\thanks{email: bunk@linde.physik.hu-berlin.de}
\\
Institut f\"ur Physik, Humboldt--Universit\"at, Invalidenstr.110,\\
10115 Berlin, Germany 
}
\date{\today}
\maketitle

\begin{abstract}
We study quenched and dynamical
pion correlators
in the two-flavour Schwinger model
(2D QED)
both
within topological charge sectors
and 
averaged.
The results
show that a clean-cut
definition
of a pion mass is no longer possible
if tunneling between sectors
does not occur.

\parag\noindent

\end{abstract}

\section{Introduction}

Constructing decorrelated configurations
in lattice simulations
can be difficult if large energy barriers 
exist 
between
regions of the configuration space
to be sampled.
One such possibility
are topological sectors
for the
$U(1)$ model.
While local updates do not work well,
we can use in two dimensions
a global heatbath update 
to obtain reasonably high tunneling rates.
This trick obviously works also in
the quenched case,
while for full dynamical simulations 
global heatbath updates 
are not known yet.

A way of
dealing with such systems
which is sometimes suggested \cite{ste97}
is to
stay in one fixed topological
charge sector and define
quantities in that way.
In order to study this for relevant observables,
we measure
pion correlators
for the massive two-flavour 
Schwinger model (QED in two dimensions) \cite{sch62}
in fixed (low)
topological sectors.
In the quenched case
we compare local and global update schemes.
For dynamical fermions
we compare low and high $\beta$ results,
using both the
local bosonic algorithm (LBA) \cite{lue94}
and Hybrid Monte Carlo (HMC) \cite{ken87}.

\section{Model}
\label{pion}

The use of the Schwinger model
with 
two degenerate flavours
of Wilson fermions
is motivated by the fact
that it is rather QCD-like,
but still tractable with 
small computational effort.
It exhibits an axial vector anomaly, light pseudoscalar mesons
(pions) and massive particles ($a_0,\eta$) \cite{het95}.

The lattice model is defined via
the action
\begin{eqnarray}
S
&=&
S_G(U)+S_F(U,\Psi),
\nn
\\
S_G
&=&
\beta \sum_P (1-{\rm Re} U_P),
\nn
\\
S_F
&=&
\sum_x \Bigl[
\bar\Psi_x \Psi_x
-
\kappa \sum_\mu
\Bigl(
\bar\Psi_x  (1+\gamma_\mu) U_{x-\mu,\mu} \Psi_{x-\mu}
+
\bar\Psi_x (1-\gamma_\mu) U^\dagger_{x,\mu} \Psi_{x+\mu} 
\Bigr) \Bigr],
\nn
\end{eqnarray}
where 
$S_G$ is the standard plaquette part
and
$S_F$
the Wilson fermion action.
$U_P$ denotes
the plaquette variable,
$U_{x,\mu}$ the compact link at site $x$ in direction $\mu$
and $\Psi$ the two-component fermion field.
We use
periodic boundary conditions for link variables
and anti-periodic ones for fermions throughout.

\section{Algorithm}

\ul{Quenched case: local update.}
We use 
a local link update 
consisting of
one exact heatbath \cite{bes79}
and three over-relaxation
steps per trajectory.

\ul{Quenched case: global update.}
Alternatively,
it is possible
to use
a global heatbath for the plaquettes.
As the new configuration
is constructed without recursion
to the old one,
we expect to encounter much less problems with metastabilities.

The global update
uses the fact that 
nearly all plaquettes are independent
even on a finite lattice.
The plaquettes have only to satisfy
the constraint
\begin{eqnarray}
\sum_P U_P  = 2\pi n; \quad n \in {\rm Z\!\!Z}.
\nn
\end{eqnarray}
We are therefore able to
update $LT-1$ plaquettes with
a heatbath algorithm
and then choose the last one  using
a Metropolis acceptance step.
This plaquette configuration
has to be translated into a valid link
configuration.
To achieve this, we utilize the freedom
to choose a gauge.
In our case we use a maximal tree prescription,
setting $LT-2$ links to 1.
Then $LT$ links can be recursively 
fixed from the plaquettes.
The unconstrained two 
remaining links
correspond to the free
Polyakov loops $P_T$ and $P_L$ 
in 2 dimensions.

We want to state
that the problem
of slow topological charge fluctuations
could also be solved
by explicit topological updates
\cite{dilger95}.
Unfortunately,
this trick is not applicable
in the presence of dynamical fermions.

\ul{Full dynamical simulations.}
In this case
we use 
the Hermitean version of
L\"uscher's local 
bosonic algorithm (LBA) \cite{lue94}.
A noisy Metropolis
step 
is used to make the algorithm exact
as
described in \cite{elser96,elser97}.
To have an independent check
we have also implemented a Hybrid Monte Carlo (HMC) algorithm.

\section{Topological sectors}
\label{topsec}

The integer-valued topological charge functional 
\begin{eqnarray}
{e\over 4 \pi} \int d^2x \epsilon_{\mu\nu} F_{\mu\nu}
\nn
\end{eqnarray}
can be represented
on the lattice
by
\begin{eqnarray}
Q 
= 
{1 \over 2\pi} \sum_{P} \phi_P
\nn
\end{eqnarray}
with plaquette angle 
$\phi_P={\rm Im} \ln (U_P) \in (-\pi,\pi)$ \cite{lue82}.
We denote by
tunnel events
all updates resulting in a 
change of the topological charge
and as tunnel probability
the number of tunnel events
divided by the total number
of updates.
The
topological susceptibility
is defined via
\begin{eqnarray}
\chi_{\rm top}
=
{1\over N_P}
[<Q^2>-<Q>^2].
\nn
\end{eqnarray}

We demonstrate the relevance
of topological sectors on a $16\times32$ lattice 
showing in Fig. \ref{tunnel} the
tunnel rate plotted
against $\beta$
for local and global updates.
\begin{figure}[tbp]
\caption{Tunnel probability as a function of $\beta$}
\label{tunnel}
\hspace{3.4cm}
\epsfxsize=7.3cm
\epsffile{tunnel_ps}
\end{figure}
The exponential decrease of the tunnel probability
for this local update algorithm
gives rise to metastabilities
in simulations at large $\beta$.
We therefore consider it worthwile
to investigate observables within fixed topological sectors.

\section{Simulations}
\label{sim}

\ul{Simulation parameters.}
We simulate on
$8\times20$ and $16\times40$
lattices
at a beta value of $\beta=12$
generating 10000 configurations.
We only show data for the larger $16\times40$ lattices.
For the fermion part,
we choose
$\kappa= 0.24$ where 
the pion correlation length (in the quenched case)
is found to be around 3
and finite size effects can be expected to be
small.

\ul{Local updates.}
We monitor the topological charge
during our simulations.
Due to metastability,
no tunneling of $Q$ is observed
in the runs
with local updates.
At $\beta = 12$ 
we therefore are able 
to perform a simulation in a given topological sector
by using an initial configuration with this particular charge.
This is done
generating a classical homogeneous
plaquette configuration
of the desired charge
and
converting this to the links
as described in Sect. \ref{pion}.
The two Polyakov loops $P_T, P_L$
are chosen according to a flat
random distribution.

\ul{Global updates.}
In the limit of independent plaquettes
the topological susceptibility
can be calculated analytically.
An approximation
for $\beta \to \infty$ 
gives
\begin{eqnarray}
\chi_{\rm top}
\approx
{1\over 4 \pi^2 \beta}.
\nn
\end{eqnarray}
which for 
$\beta=12$
amounts to $
\chi_{\rm top}
=
0.0022 $.
To check our global update,
we simulate
on a $16 \times 32$ lattice
obtaining
$
\chi_{\rm top}
=
0.0021(1)
$.
This shows that the global update works well
without metastabilities in the topological charge.

\ul{Observables.}
We measure the pion correlator 
using the operator
$ \bar\psi \gamma_5 \tau \psi$.
We apply a point source
and wall sink prescription
to increase statistics \cite{elser96}.
The effective masses
are calculated from two adjacent time slices
using
the hyperbolical cosine formula.
The error analysis of the effective pion mass
takes covariances and auto-correlations of the correlators
into account.

\subsection{Quenched results}

\begin{figure}[tbp]
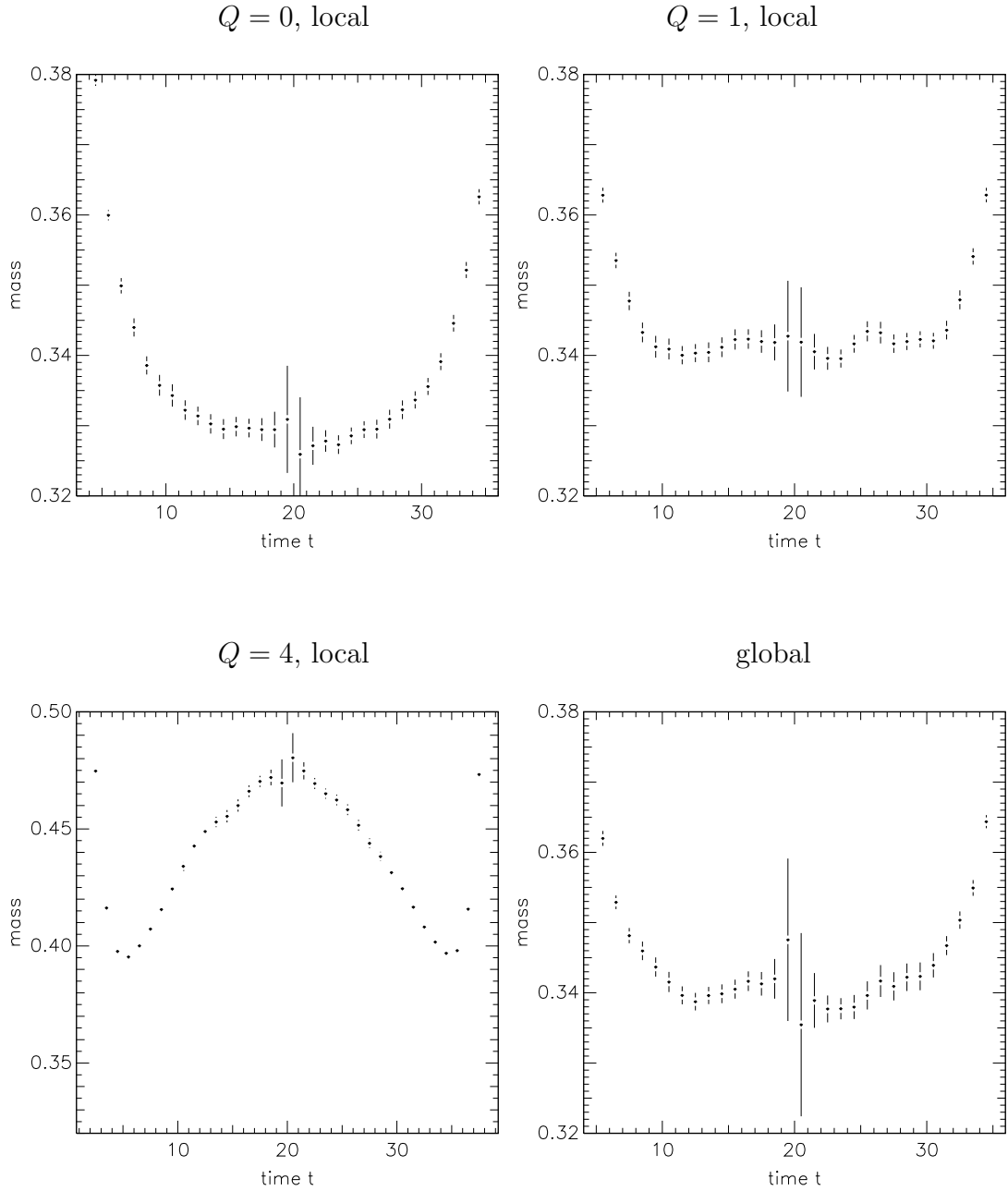

\caption{Quenched effective pion mass 
as a function of time for $\beta=12.0$, $\kappa=0.24$}
\label{quenched12}
\unitlength1cm

\vspace{0.5cm}
\hspace{3cm}
$Q=0$, local
\hspace{4cm}
$Q=1$, local
\vspace{-0.5cm}

\begin{picture}(20,8)
\epsfxsize=7.3cm
\epsffile{p1_ps}
\epsfxsize=7.3cm
\epsffile{p2_ps}
\end{picture}

\vspace{1cm}
\hspace{3cm}
$Q=4$, local
\hspace{5cm}
global
\vspace{-0.5cm}

\begin{picture}(20,8)
\epsfxsize=7.3cm
\epsffile{p3_ps}
\epsfxsize=7.3cm
\epsffile{p5_ps}
\end{picture}

\end{figure}

The results of the quenched runs
are shown in Fig. \ref{quenched12}.
\begin{figure}[tbp]
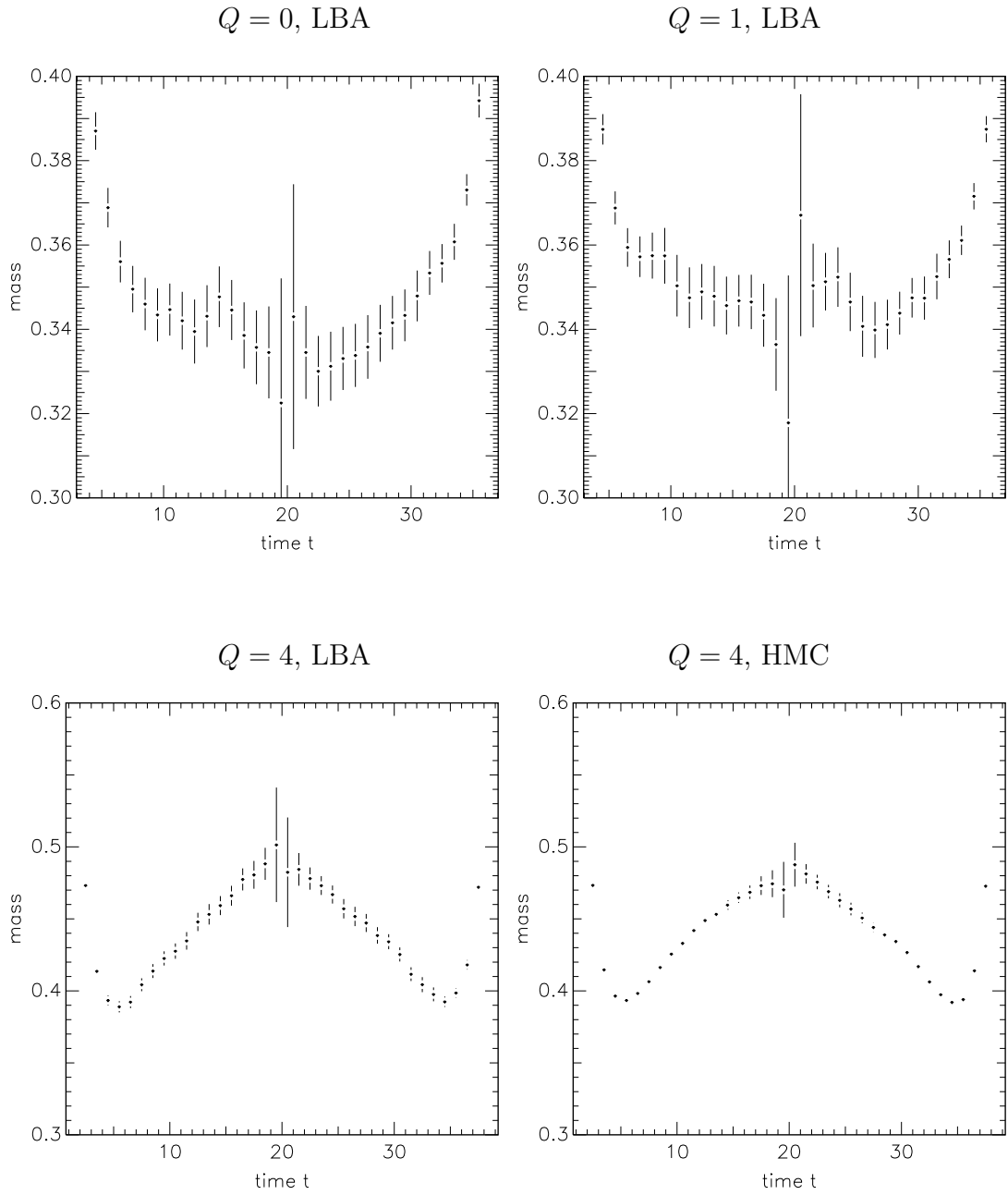

\caption{Dynamical effective pion mass as a function of time for 
$\beta=12.0$, $\kappa=0.24$}
\label{dyn12}
\unitlength1cm

\vspace{0.5cm}
\hspace{3cm}
$Q=0$, LBA
\hspace{4cm}
$Q=1$, LBA
\vspace{-0.5cm}

\begin{picture}(20,8)
\epsfxsize=7.3cm
\epsffile{p6_ps}
\epsfxsize=7.3cm
\epsffile{p7_ps}
\end{picture}

\vspace{1cm}
\hspace{3cm}
$Q=4$, LBA
\hspace{4cm}
$Q=4$, HMC
\vspace{-0.5cm}

\begin{picture}(20,8)
\epsfxsize=7.3cm
\epsffile{p8_ps}
\epsfxsize=7.3cm
\epsffile{p10_ps}
\end{picture}

\end{figure}
Obviously,
for $Q=0,4$ we are not able to find
a plateau in the effective mass plots.
We find a valley-like structure
in the mass
for the low $Q$ cases.
For high $Q$
this turns into 
a hill-like structure with the peak situated at
half of the temporal lattice extent.
To show 
that in the intermediate region
the valley and hill structures can approximately cancel
and suggest a fake plateau,
we 
include the $Q=1$ plot.

In the quenched case,
we are able to compare to the
results using a global update scheme.
For the global update we find a tunnel probability
of $P=0.76$ and therefore do not
expect any influence of topology.
The effective mass
is shown in Fig. \ref{quenched12}d.
A plateau is clearly 
more reasonable than in the fixed $Q$ cases.

\subsection{Dynamical fermion results}

Effective masses from
full dynamical simulations
are 
shown in Fig. \ref{dyn12}.
We do not expect quantitatively the same results
as in the quenched case.
But the behaviour is
qualitatively similar.

To establish
that our dynamical results
are not influenced by
the chosen parameters of the local bosonic algorithm,
we repeat the calculation
with topological charge $Q=4$
using a standard HMC algorithm.
This is also shown in Fig. \ref{dyn12}.
The results agree nicely within errorbars.

From these results,
we conclude
that we need 
to average over the topological sectors
to obtain a plateau in the effective mass.

\section{Projections to topological sectors}
\label{project}

To gain further insight,
we now use a slightly different approach.
In principle, we
could also restrict ourselves
to definite topological sectors
by selecting measurements
with fixed topological charge 
from a 
simulation,
i.e. effectively simulating the path integral
given by
\begin{eqnarray}
Z[q] = \int D[U] D[\bar\Psi] D[\Psi] \delta_{Q,q} e^{-S}.
\nn
\end{eqnarray}
To this end
we need simulations with a reasonably high fluctuation.
Such simulations can be done e.g. 
with dynamical simulations at low $\beta$
or quenched simulations using global updates.

\begin{figure}[tbp]
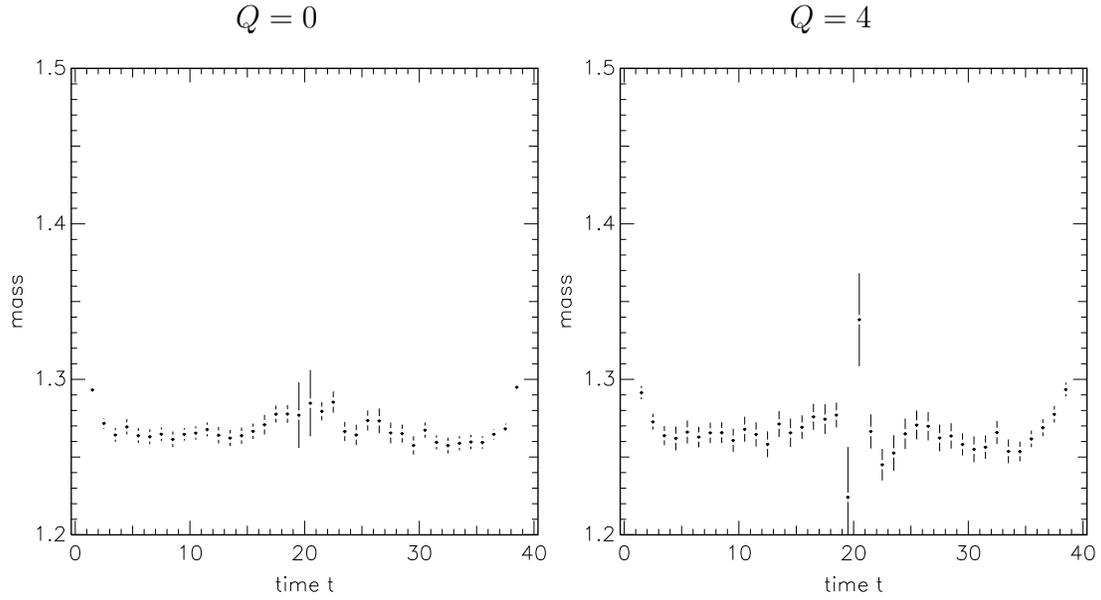

\caption{Dynamical projected effective pion mass
as a function of time for
$\beta=1.0$, $\kappa=0.22$}
\label{dyn1-fixedQ}
\unitlength1cm

\vspace{0.5cm}
\hspace{3cm}
$Q=0$
\hspace{6cm}
$Q=4$
\vspace{-0.5cm}

\begin{picture}(20,8)
\epsfxsize=7.3cm
\epsffile{p-dyn1-fixedQ_1_ps}
\epsfxsize=7.3cm
\epsffile{p-dyn1-fixedQ_2_ps}
\end{picture}

\end{figure}
\begin{figure}[tbp]
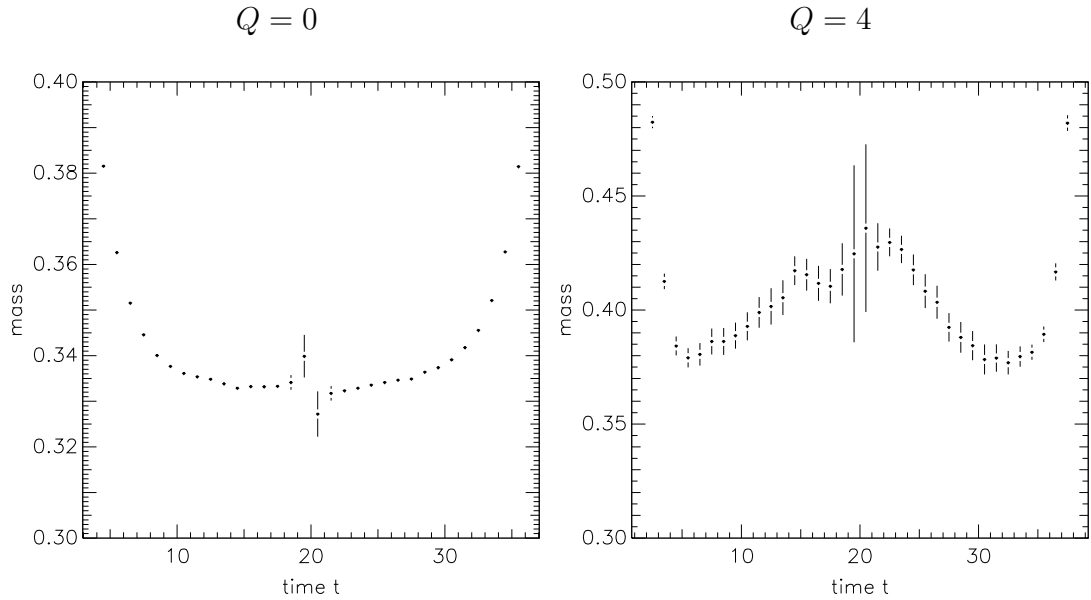

\caption{Quenched projected effective pion mass
as a function of time for 
 $\beta=12.0$, $\kappa=0.24$}
\label{q12-fixedQ}
\unitlength1cm

\vspace{0.5cm}
\hspace{3cm}
$Q=0$
\hspace{6cm}
$Q=4$
\vspace{-0.5cm}

\begin{picture}(20,8)
\epsfxsize=7.3cm
\epsffile{p-q12-fixedQ_1_ps}
\epsfxsize=7.3cm
\epsffile{p-q12-fixedQ_2_ps}
\end{picture}

\end{figure}

\ul{Full dynamical case.}
We work at low $\beta=1$
with a slightly smaller $\kappa=0.22$.
Effective masses are depicted in Fig.
 \ref{dyn1-fixedQ}.
We can detect no discrepancy between
masses calculated in different topological sectors.
This result
was also reported by a group working with staggered fermions,
which concluded that topological sectors to not matter
for the Schwinger model
\cite{ste97}.

\ul{Quenched case.}
Here we exploit the opportunity
to use the same parameters $\beta=12,\kappa=0.24$
as in Sect. \ref{sim}.
The results are shown in  Fig.
 \ref{q12-fixedQ}.
At this $\beta$, we do not find 
agreement.
Rather the effective masses
are
nearly the same 
as in the simulations without tunneling
presented in Fig. \ref{dyn12}.
We would like to point out
that they do not agree within errorbars.
On the other hand, we remark that the statistical sample
was very much smaller for the projected data
due to the fact that only a part of the generated
configurations
is projected into
the appropriate sectors.

The striking difference between low and high $\beta$ results
makes a sound understanding of this
behaviour
highly desirable.
It is evidently not just the
averaging over the topological sectors (as found in Sect. \ref{sim})
which is lacking in high $\beta$ simulations,
but there seems to be some more subtle dynamical 
effect involved.

\section{External gauge configurations}
\label{external}

\begin{figure}[tbp]
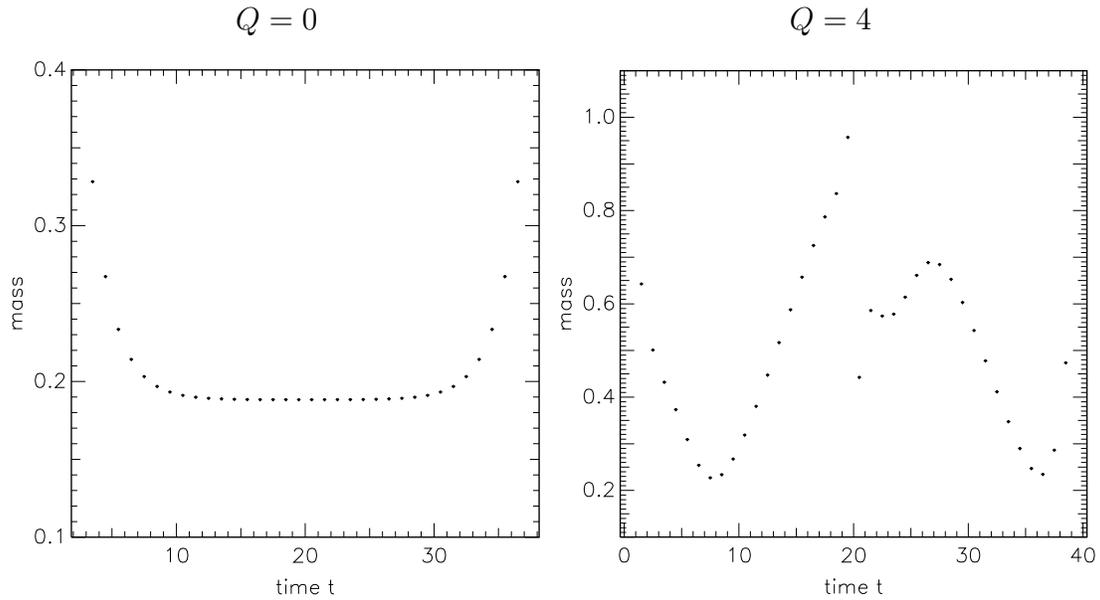

\caption{Effective pion mass 
as a function of time 
from one external configuration at $\kappa=0.24$ }
\label{config1}
\unitlength1cm

\vspace{0.5cm}
\hspace{3cm}
$Q=0$
\hspace{6cm}
$Q=4$
\vspace{-0.5cm}

\begin{picture}(20,8)
\epsfxsize=7.3cm
\epsffile{p-config1_1_ps}
\epsfxsize=7.3cm
\epsffile{p-config1_2_ps}
\end{picture}

\end{figure}

In order to further investigate 
how much averaging is necessary
we plot in Fig. \ref{config1} effective 
masses for external
configurations with 
fixed topological charge
$Q=0$ and $Q=4$.
These are generated
from homogeneous
plaquette configurations
in the way described in Sect. \ref{pion}.
For the fermions we use
$\kappa=0.24$. 

We stress that we use random Polyakov loops $P_T, P_L$
in both cases,
so that one should not expect free fermion behaviour in the $Q=0$ case.
For $Q=4$ a completely irregular behaviour is observed. 
It is thus not possible to measure
a meaningful pion correlator from one (even very smooth)
configuration alone. The translation invariance is
manifestly broken
even for that smooth configuration.

\begin{figure}[tbp]
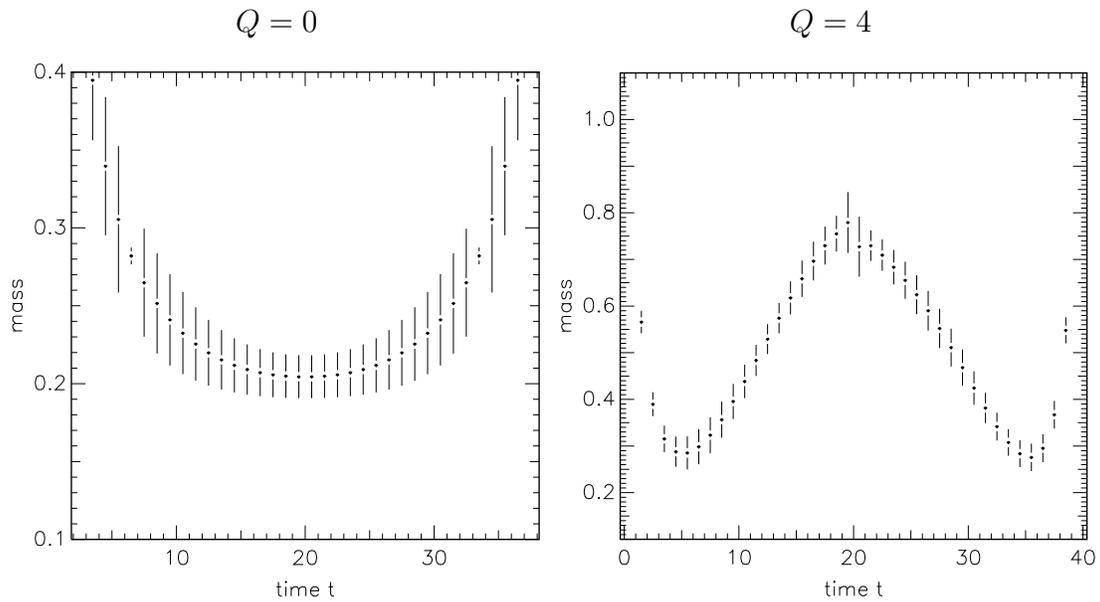

\caption{Effective pion mass 
as a function of time 
from ten external configurations at $\kappa=0.24$}
\label{config10}
\unitlength1cm

\vspace{0.5cm}
\hspace{3cm}
$Q=0$
\hspace{6cm}
$Q=4$
\vspace{-0.5cm}

\begin{picture}(20,8)
\epsfxsize=7.3cm
\epsffile{p-config10_1_ps}
\epsfxsize=7.3cm
\epsffile{p-config10_2_ps}
\end{picture}

\end{figure}

The result of 
averaging over 10 values of  $P_T$ and $P_L$
is shown in Fig. \ref{config10}.
We clearly regain the qualitatively
expected regular valley and hill structure 
observed in Fig. \ref{dyn12}
in both the
$Q=0$ and $Q=4$ case.

\section{Conclusion}

Our results for effective pion masses
for simulations at fixed topological charge
at high $\beta$
clearly show that
definition of a 
pion mass 
from a plateau is not possible 
for both quenched and dynamical simulations
even for vanishing topological charge.

A comparison with quenched 
global updates exhibiting no metastabilities
demonstrates
that a plateau can be found
in the correct path integral sample.
Furthermore, 
cross-checks against HMC
indicate that this problem
is not an artifact
stemming from the fermion algorithm.

In Sect. \ref{project},
we studied
effective pion masses
calculated from 
projections
to fixed topological sectors. 
Results
from
fluctuating
ensembles
show no dependence on the
topological charge.
On the other hand,
those projected from non-fluctuating
ensembles
show approximately
the same behaviour
as simulations completely without tunneling.
This suggests a rather subtle dynamical
effect.

In Sect. \ref{external}
we found that
external homogeneous plaquette configurations
with 
fixed Polyakov loop values $P_T$ and $P_L$
exhibit completly irregular behaviour.
After averaging over $P_T$ and $P_L$
we regain the 
effective mass results
characteristic for the
topological charge sectors these 
configurations lie in.

We conclude that
there is a 
need to obtain a better understanding
of the interplay between
the dynamical mass generation and 
topological sectors.
We would like to point out that
higher statistics runs
could reveal similar phenomena
in other models with non-trivial topological structure.

\section{Acknowledgements}
We
would like to thank W. Bock for discussions.
Part of the calculations were
done
on the Fujitsu VPP500 provided by
ZIB Berlin and the IBM SP2 at
Desy--IfH Zeuthen.


\end{document}